\documentclass{iau}
\usepackage{graphicx} 

\title[IAUS291.~~Complex pulsar magnetosphere] %% short title %%
{The complex charm of the pulsar magnetosphere} %% full title %%

\author[A.~N.~Timokhin]  %% short author list %%
{A.~N.~Timokhin
}

\affiliation{Astrophysics Science Division, NASA Goddard Space
  Flight Center,
  Greenbelt,  USA\\[\affilskip]
  Moscow State University, Sternberg Astronomical Institute, Moscow,
  Russia
\\[\affilskip]

email: {\tt andrey.timokhin@nasa.gov} }

\pubyear{2012}
\volume{291}  %% insert here IAU Symposium No.
\jname{\mbox{Neutron Stars and Pulsars: Challenges and Opportunities after 80 years}}
\editors{J. van Leeuwen, ed.}

\newcommand{\GJ}[1]{\ensuremath{#1_{\textrm{\tiny{}GJ}}}}    
\newcommand{\jm}{\ensuremath{j_{\mathrm{m}}}}

\newcommand{\mnras}{\textit{MNRAS}}
\newcommand{\apj}{\textit{ApJ}}
\newcommand{\apjl}{\textit{ApJ Letters}}
\newcommand{\aap}{\textit{A\&A}}

\begin{document}

\maketitle

%% -- Abstract ----------------------------------
\begin{abstract}
  I give a brief overview of recent results from self-consistent
  modeling of electron-positrons cascades in pulsar polar caps.  These
  results strongly suggest that the pulsar magnetosphere is a more
  complex system than was assumed before.
\end{abstract}

% add below any authors, subjects and objects for indexing 
%   add more lines if necessary
%   but leave all lines commented out
%\index[author]{LastName1, Initials|textbf}
%\index[author]{LastName2, Initials|textbf}
%\index[subject]{Keyword1}
%\index[subject]{Keyword2}
%\index[object]{Object1}
%\index[object]{Object2}

\firstsection % if your document starts with a section,
\section{Introduction}

Radio pulsars are rotationally powered highly magnetized isolated
neutrons stars (NS) which emission is produced in their
magnetospheres.  There are strong observational evidences that pulsar
magnetospheres are filled with dense plasma -- pulsar wind nebulae
(PWNe) are fed by dense flow of relativistic plasma produced by their
parent pulsars. Most of theoretical models of pulsar magnetospheres,
starting from the classical model of Goldreich \& Julian (1969), also
argue that pulsar magnetosphere is filled with plasma.  The sharpness
of peaks in pulsar light curves, especially in gamma, naturally leads
to the conclusion that emitting regions, and, hence, the regions where
particles are being accelerated, are small and everywhere in the
magnetosphere, except those small accelerating regions, the electric
field is screened by dense plasma.

The vast majority of NS energy losses goes into the pulsar wind, the
outflow of relativistic plasma. Motion of the plasma in the
magnetosphere results in electric currents which change the topology
of the NS's magnetic field creating closed and open magnetic field
lines zones. Plasma flow along open magnetic field lines starts at the
NS surface and leaves the magnetosphere; that plasma must be
constantly replenished.  Even if charged particles can be extracted
from NS surface, their number density would be several orders of
magnitude lower than that inferred from observations of PWNe. Starting
with the work of Sturrock (1971) the assumption about electron-positron
plasma generation in pulsar polar caps has been an integral part of
almost any pulsar model.  It is also generally believed that pair
creation is intimately connected to radio pulsar activity, as the
death line -- the place on the $P-\dot{P}$ diagram where radio
emission ceases -- roughly corresponds to such pulsar parameters when
the potential drop generated by NS rotation becomes smaller than the
threshold for pair formation. Hence, production of electron-positron
plasma in polar caps is a cornerstone of current pulsar ``standard
model''.

The problem of how plasma is generated in pulsar polar caps can not be
considered separately from the problem of the global structure of the
magnetosphere.  Currents supporting the magnetosphere with its open
and closed field lines zones flow along magnetic field lines all the
way from the NS surface into the pulsar wind zone passing trough the
plasma generating regions.  Current density distribution is determined
by the global magnetospheric structure and those small pair generation
zones (which inductance is much smaller than that of the
magnetosphere) must adjust to the current density imposed by the
magnetosphere.  Recently significant progress has been achieved in
modeling of the global structure of pulsar magnetosphere
(e.g. \cite{CKF,Timokhin2006,Spitkovsky2006,Kalapotharakos2009}), so
the current density distribution in the magnetosphere is known.  It
has been explicitly shown (\cite{Timokhin2006}) that this current
density distribution does not agree with assumptions about the current
density used in than up to date quantitative ``standard models'' of
polar cap plasma generation
(e.g.~\cite{AronsScharlemann1979,MuslimovTsygan1992,DaughertyHarding1982}),
which assumed stationary unidirectional plasma flow.

This discrepancy motivated me to start the study of pair plasma
generation in pulsar polar caps which is free from assumptions about
character of plasma flow and addresses the problem starting from first
principles. The goal was to investigates how the pair plasma is
generated when a given current density (set by the global
magnetosphere structure) flows through the pair creating region.  Here
I give a brief overview of the first results of this study, described
in more detail in Timokhin (2010) and Timokhin \& Arons (2012), and
discuss possible explanations for several phenomenas seen in pulsars.

\section{Self-consistent numerical model of pair cascades}

We assumed that pulsar magnetosphere is already filled with plasma and
studied how this state is sustained%
\footnote{In other words, we did not study the (more difficult) problem
  of how the magnetosphere is formed.}. %
In the reference frame corotating with the NS, the star's rotation results
in the effective background charge density, the Goldreich-Julian (GJ)
charge density $\GJ{\eta}$.  Existence of the open magnetic field
lines requires these lines to be twisted; this twist must be supported
by a certain current density $\jm$ which flows along the lines and
through the cascade zone as well. Both these effects must be included
in modeling of electrodynamics of the cascade zone, but almost all
previous quantitative models of pair cascades did not include the
inductive effects, i.e. they ignored $\jm$.  We modeled how the
cascade zone behaves under different current loads -- in each
simulation we required that a given current density $\jm$ flows
through the cascade zone; in contrast to almost all previous works we
studies the pair production when the current is fixed rather that the
voltage.

We used a specially developed hybrid Particle-In-Cell/Monte Carlo
(PIC/MC) numerical code which models electromagnetically driven pair
cascades in truly self-consistent way, whereby particle acceleration,
photon emission, propagation, pair creation, and screening of the
electric field are calculated simultaneously (Timokhin 2009, 2010).
As such truly self-consistent simulations had never been done before
we started with the simplest possible model which, however, includes
all types of physical processes relevant for pair formation in the
polar caps of pulsars.  Our model is one-dimensional, it includes
curvature radiation as a gamma-ray emission mechanism and single
photon absorption in strong magnetic field as a pair production
mechanism.  The electrodynamics takes into account the effects due to
the GJ charge density $\GJ{\eta}$ as well as the current density $\jm$
imposed by the magnetosphere. The electrodynamics and plasma dynamics
-- particle acceleration and electric field screening by charged
particles -- are modeled by the PIC part of the code. Emission of
gamma-rays, their propagation in magnetic field, and pair creation are
modeled by the MC part of the code.

Boundary conditions implemented in the code included the case when
particles cannot leave the NS surface as well as the case of free
particle outflow from the surface, the so-called Space Charge Limited
Flow (SCLF) regime.  The latter, less trivial case was modeled by
creation of a pool of numerical particles just outside of
computational domain at its NS's end. The system was allowed to
extract as many particles as it needed, in other words we allowed the
cascade zone to set the electric field at the NS to zero
self-consistently, without imposing it in the code manually. Particles
were allowed to leave domain freely (if not prevented to do so by the
electric field) and no particles were injected at the outer end of the
domain.

\section{Main results}

We performed self-consistent simulations of pair cascades in pulsar
polar cap in 1D for two most important classes of pulsar polar cap
cascade models (i) when particles cannot be extracted from the NS
surface (\cite{Timokhin2010a}), the so-called Ruderman-Sutherland 
(1975; hereafter RS) model; and (ii) for currently the most popular model when
particles can freely leave the surface (\cite{TimokhinArons2012}), the
space charge limited flow regime, so-called Arons-Scharlemann (1979)
model.

In both cases the cascade zone easily adjusts to \emph{any} given
current density $\jm$ imposed by the magnetosphere provided the
physical parameters allow for pair creation. This adjustments proceeds
locally due trapping of some fraction of plasma particles by small
fluctuating electric field.  $\jm$ turned out to be the most important
parameter determining the efficiency of particle acceleration.  In
some cases sustaining if the imposed current density results in a flow
with no particles acceleration and pair creation.  If the imposed
current density leads to pair formation, it always occurs
non-stationary, a burst of pair formation is followed by a quiet phase
when accelerating electric field is screened and no pairs are
produced.

For the Ruderman-Sutherland model the cascade easily adjusts to the
current density required by the magnetosphere and always produces
dense electron-positron plasma in accordance with qualitative
expectations of the original model, provided $\jm\neq0$.  Particle
acceleration and pair production occur in form of discharges.  At the
beginning of each discharge cycle a gap (a charge starved spatial
region) with accelerating electric field appears and grows in size
until the potential drop across it becomes larger than the pair
formation threshold. Particles accelerated in this gap emit pair
production capable gamma-rays which inject electrons and positron into
the gap, these secondary particles screen the electric field and
destroy the gap. When the newly generated plasma leaves the domain
the discharge starts anew.  Surprisingly, the pair formation turned
out to be very regular showing a limit cycle behavior, and gaps do not
stay at the same place but move along magnetic field lines.  The pair
plasma has a thermalized low-energy component.

% CUP work flow only accepts EPS -- not PDF, JPG, etc.
\begin{figure}
\begin{center}
 \includegraphics[width=\textwidth]{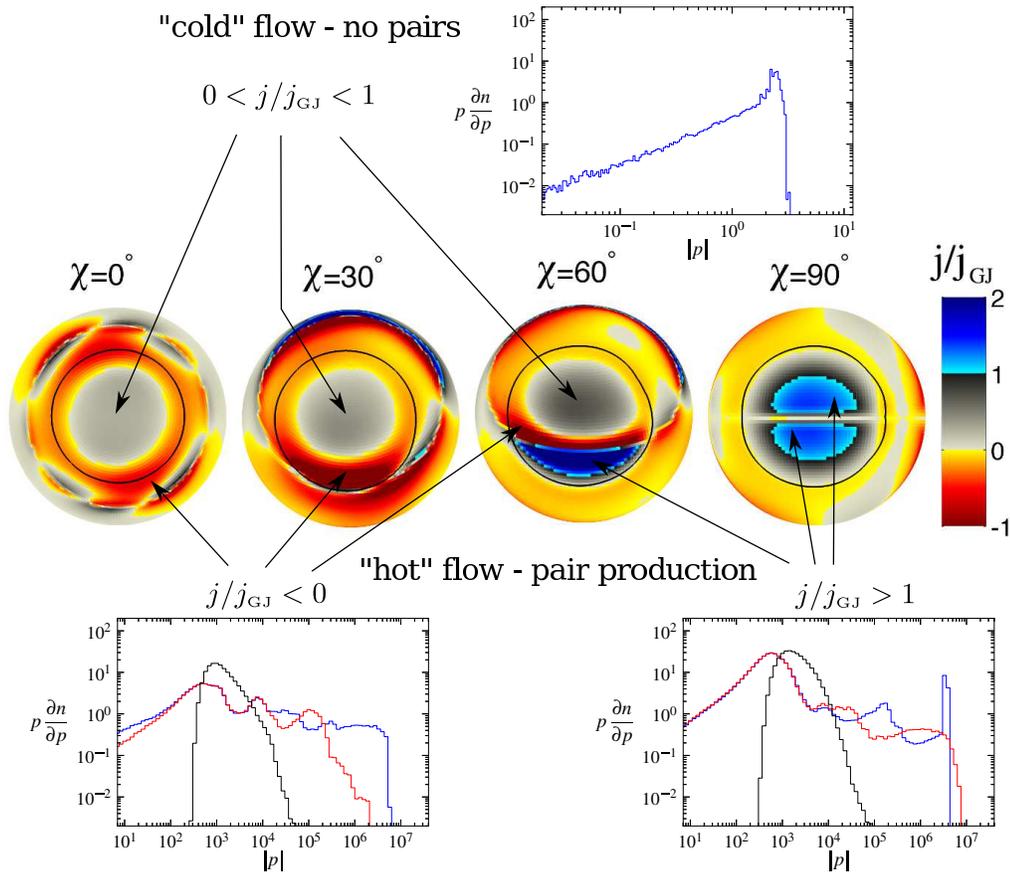} 
 \caption{Current density distribution in the polar cap of pulsar for
   different pulsar inclination angles $\alpha$ (central panel) and
   examples of particle distribution functions (as functions of
   particle momenta normalized to $m_ec$) for different cascade
   regimes in the space-charge limited flow model (top and bottom
   panels).  Colors show the ratio of $j/\GJ{j}$ and the polar cap
   boundary is shown by a thin black circle on each subplot.
   Distribution function of electrons is shown by blue lines,
   positrons by red dashed lines, and gamma-rays by dotted black
   lines.  Note that on the upper plot particles are only mildly
   relativistic and no pairs are produced (adapted from Timokhin \&
   Arons (2012) with contribution of Xue-Ning Bai 
   (\cite{BaiSpitkovsky2010a}))}
   \label{fig:j_vs_seds}
\end{center}
\end{figure}

In the case of the space charge limited flow, however, the cascade
behavior turned out to be \emph{qualitatively} different from what was
expected in ``standard'' cascade models.  The character of the flow
strongly depends on the ratio of the average current density flowing
through the cascade zone to the GJ current density
$\GJ{j}\equiv\GJ{\eta}c$, see Fig.~\ref{fig:j_vs_seds}.  For field
lines where the imposed current density is smaller that the GJ current
density $0<\jm/\GJ{j}<1$ (sub-GJ) no pair plasma is produced%
\footnote{in this regard our results support conjectures about the
  sub-GJ flow of Shibata (1997) and Beloborodov (2008)} %
because the accelerating zone is very small due to an instability of
the plasma flow and a moderately relativistic electron low-density
plasma (with the number density $n=\GJ{\eta}/e$) streams along those
field lines.  Pair formation is possible only along field lines where
the current density is either larger than the GJ current density
$\jm/\GJ{j}>1$ (super-GJ), or has the opposite sign to it
$\jm/\GJ{j}<0$, in regions with the return current.  Pair creation is
highly non-stationary, similar to discharges in the RS model.  SCLF
regime can sustain any imposed current density $\jm$ as well.

Contrary to expectations of previous models, the place where
discharges occur is different for different flow regimes. For RS
cascades with $\jm/\GJ{j}>0$ discharges start close to the NS
surface. For flows with $\jm/\GJ{j}<0$ discharges start at the largest
possible distance from the NS in both RS and SCLF regimes. For SCLF
with $\jm/\GJ{j}>1$ the position where discharges start depend on how
the GJ charge density changes with the distance: discharges can start
close to NS if the ratio $|\GJ{\eta}/B|$ ($B$ -- magnetic field
strength) increases with the distance from the NS, otherwise
discharges start at large distances from the NS.

\begin{figure}
\begin{center}
 \includegraphics[width=\textwidth]{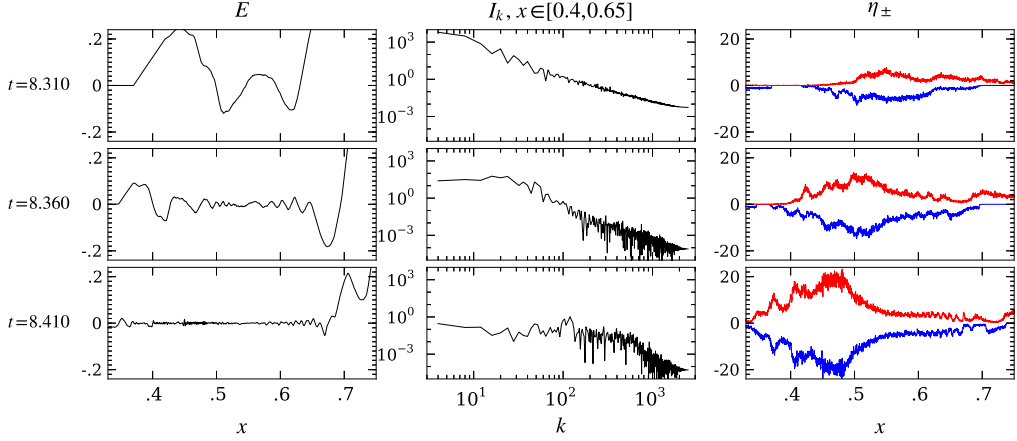} 
 \caption{Screening of the electric field and formation of
   superluminal electrostatic waves during a discharge for SCLF with
   $\jm=-0.5\GJ{j}$ (cf Fig.~19 in Timokhin \& Arons 2012).  There are
   three snapshots for the electric field $E$, power spectra of the
   electric field $I_k=|E_k|^2$ (for the spatial interval
   $x\in[.4,.65]$), and the charge density of electrons (negative
   values, blue line) and positrons (positive values, red line)
   $\eta_{\pm}$. $E$ and $\eta_{\pm}$ are plotted as functions of
   distance $x$ for the part of the calculation domain with intense
   pair formation.  $x$ is normalized to the domain size $L$, $E$
   normalized to the ``vacuum'' electric field
   $E_0\equiv|\GJ{\eta}|\pi{}L$, $\eta_{\pm}$ is normalized to the
   absolute value of the Goldreich-Julian charge density
   $|\GJ{\eta}|$. Time $t$ is measured in flyby time $L/c$.}
   \label{fig:wave_discharge}
\end{center}
\end{figure}

Discharges results in strongly fluctuating electric field,
electrostatic waves.  Fig.~\ref{fig:wave_discharge} shows an example
of how the screening of the electric field in a discharge proceeds,
there are 3 snapshots of a discharge in SCLF with $\jm=-0.5\GJ{j}$.
Fluctuating electric field during discharge event has a power low
spectrum, with long-wavelength (small $k$) cut-off moving to larger
$k$ as the wavelength of fluctuations decreases.  The phase velocity
of such waves is larger than the light speed and they can
not be effectively dumped via Landau damping.

\section{Discussion}

Results of these simulations imply that the pulsar magnetosphere is a
much more complex physical system than it was assumed before.  For the
same pulsar period and magnetic field strength properties of plasma
flowing along a given magnetic field line strongly depend on the value
of the imposed current density $\jm$ along that line.  Plasma
properties (density, particle energy distribution) along different
magnetic field lines can differs substantially due to non-uniform
distribution of $\jm$ across the polar cap, and plasma content of
magnetospheres in pulsars with different inclination angles will also
differ as the current density distribution $\jm$ strongly depends on
the inclination angle (see the middle panel of Fig.~\ref{fig:j_vs_seds}).

The locations of particle acceleration and emission zones depend in a
non-trivial way on the pulsar inclination angle.  For example, in the SCLF
regime there is no pair plasma generation over the most areas of the
polar cap in an aligned pulsar, but in an orthogonal rotator pair
plasma is efficiently generated over the whole polar cap.  Our results
also indicate that magnetic field lines with the return current
($\jm/\GJ{j}<0$) can have particle acceleration zones in the outer
magnetosphere, as discharges tend to start at the furthest possible
distance from the NS.  This agrees with observations of pulsars with
\textit{Fermi} which indicate that gamma-rays are produced in the
outer magnetosphere, in regions close to those where the field lines
carrying the return current are expected to be.

Non-stationary discharges in flow regimes with pair creation
incorporate time dependent, quasi-coherent currents on microsecond and
shorter time scales.  Such fluctuations might be a direct source of
radio emission from the low altitude polar flux tube, a region
strongly suggested as the site of the radio emission by the radio
astronomical phenomenology. The energy in such fluctuations is enough
to power the radio emission and the spectrum of the fluctuations is a
power law, consistent with radio phenomenology.  Fluctuating electric
field is also present in the domain the low energy flow with sub-GJ
current density $0<\jm/\GJ{j}<1$ in SCLF regime, however, the
amplitude of this field is so low that it is unlikely that those
fluctuations could directly result in observable radio emission.

It is natural to assume that all these different flow regimes have
different observational signatures, i.e. are responsible for different
components in pulse profiles.  If so, from the current density
distribution (the central panel on Fig.~\ref{fig:j_vs_seds}) one can
see that pulsar profiles should be roughly symmetric and the maximum
number of separate emission regions should not exceeds 5, what seems
to agree with results of phenomenological analysis of pulsar profiles
(\cite{Rankin1983}).

Changes in $\jm$ could result in significant changes of pulsar
emission.  For example, in SCLF regime changing $\jm$ from super-GJ to
sub-GJ will result in highly relativistic plasma flow becoming a low
energetic one.  If pulsar magnetosphere has a few metastable states
with different current density distributions, then the character of
radio emission could be qualitatively different in these two states;
it could be that there will be no radio emission in one of the states
state at all. This might be a low-level mechanism for nulling and/or
mode changing in (at least some of) pulsars.

It must be said, however, that the resulting 1D model of the cascades
is very simplified.  Within the frame of 1D model many important
issues cannot be addressed, such as influence of physical conditions
at adjacent field lines on the accelerating electric field and
excitation and propagation of electromagnetic waves.  In SCLF regime
the spatial scales involved are larger that the polar cap size, the
characteristic transverse size of the system, what makes 1D model not
suitable for accurate quantitative predictions.  However, we expect
that most of qualitative results obtained with the current model holds
a multi-D treatment which will be reported in later papers.

~\\
{\bf Acknowledgements:}
I was supported by an appointment to the NASA Postdoctoral program at
NASA Goddard Space Flight center, administered by ORAU, and by the
\textit{Fermi} Guest Investigator Program

\end{document}